\documentclass[preprint,aps,prd,twocolumn,10pt]{revtex4}

\usepackage{amsmath,amssymb}

\usepackage{color}

\usepackage[dvips]{graphics}
\usepackage{rotating}
\usepackage{epsfig}
\usepackage{amsmath}
\def\lesssim{\mathrel{\hbox{\rlap{\hbox{\lower5pt\hbox{$\sim$}}}\hbox{$<$}}}}
\def\gtrsim{\mathrel{\hbox{\rlap{\hbox{\lower5pt\hbox{$\sim$}}}\hbox{$>$}}}}

\begin{document}

\preprint{KIAS-P11070, HIP-2011-33/TH}

\title
{Probing Higgs in Type III Seesaw at the LHC}

\author{Priyotosh Bandyopadhyay$^{a,c}$,  Suyong Choi$^b$, and Eung Jin
Chun$^{a}$,  Kyungnam Min$^{b}$}

\address{$^a$Korea Institute for Advanced Study, Seoul 130-722, Korea\\
$^b$Department of Physics, Korea University, Seoul 136-713, Korea\\
$^c$Department of Physics  and
    Helsinki Institute of Physics, University of Helsinki,
  FIN-00014,  Finland\\
}


\begin{abstract}
{We show that the type III seesaw mechanism opens up a promising
possibility of searching the Higgs boson in the $b \bar b$ channel
through the Higgs production associated with a charged lepton
coming from the decay of the triplet seesaw particle. In
particular we look for the $2b$ signals with trileptons or
same-sign dileptons to construct the Higgs and the triplet fermion
mass and calculate the reach with the integrated luminosity of 10
fb$^{-1}$ at the 14 TeV LHC}.
\end{abstract}


\maketitle


Finding the Higgs boson and thus verifying the electroweak
symmetry breaking mechanism is a primary goal of the LHC. 
The recent LHC data constrain the Higgs mass in a narrow range of 115--130 GeV \cite{lhc1213}.
For such a low mass Higgs boson, its discovery relies on the combination of several
channels based on the gluon fusion production, in particular, $gg \to h\to \gamma\gamma$.
Another interesting channel is the associated Higgs boson production
$pp \to W(Z) h$ followed by the dominant Higgs decay $h\to b\bar b$
and leptonic decays of $W(Z)$ for which a low significance due to large backgrounds
can be overcome by using subjet techniques in a boosted regime \cite{Butterworth08}.
Probing such a channel is important as it can provide an independent information
on the Higgs boson coupling to gauge bosons and $b$ quarks.

The purpose of this work is to explore another possibility for the
Higgs discovery which arises in the type III seesaw mechanism, 
 where the origin of the observed neutrino masses and mixing
is attributed to $SU(2)$ triplet fermions with hypercharge zero
\cite{Foot88}. Such new particles can be produced through the electroweak
interaction and subsequently decay to a lepton plus $W$, $Z$ or
$h$. These signatures can be traced successfully to reconstruct
the new triplets and thus confirm the type III seesaw mechanism
\cite{Hambye08,Aguila08,Arhrib09,Bandyo09,Li09}. Among these type
III seesaw signatures, we focus on the Higgs production associated
with a charged lepton followed by the Higgs decay $h\to b\bar b$
to show that this channel provides a promising search channel for
the low mass Higgs boson.

\smallskip

The type III seesaw mechanism introduces $SU(2)_L$ triplet fermions with $Y=0$,
$\Sigma=(\Sigma^+, \Sigma^0, \Sigma^-)$, which can be written in a matrix form:
 \begin{equation}
 {\bf \Sigma} = \begin{pmatrix}
      \Sigma^0 & \sqrt{2} \Sigma^+ \cr
      \sqrt{2} \Sigma^- & -\Sigma^0
       \end{pmatrix} .
 \end{equation}
Then the gauge invariant Yukawa terms are
 \begin{equation}
 {\cal L} = \left[
 y_i H \varepsilon  \bar{\bf \Sigma} P_L l_i  + h.c. \right]
  +{1\over4} m_\Sigma \mbox{Tr}\left[ \bar{{\bf\Sigma}}{\bf\Sigma}\right]
 \end{equation}
where $l_i$ is the lepton doublet and $H$ is the Higgs doublet:
$l_i=(\nu_i, e_i)_L$ and $H=(H^+, H^0)$.  In the unitary gauge,
$H^+=0$ and $H^0 = v+h/\sqrt{2}$ with $v=174$ GeV, we get
\begin{equation}
 {\cal L}_{Yukawa} = y_i \left[ \sqrt{2} \bar{\Sigma}^- P_L e_i +
  \bar{\Sigma}^0 P_L \nu_i + h.c.\right]
  {h\over\sqrt{2}}  \,.
 \end{equation}
Here we take only one generation of the triplet for our
illustration. Note that the neutrinos get a seesaw mass  $m^\nu_{
ij}= y_i y_j v^2/m_\Sigma$ which becomes of the order 0.1 eV for $y_i
\sim 10^{-6}$ and $m_\Sigma \sim 1$ TeV.  The neutrino Dirac mass, $y_i
v$, induces mixing between $l$ and $\Sigma$. The mixing angles for
the neutral and charged part are
 \begin{equation}
 \theta_{\nu_i} \approx {y_i v\over m_\Sigma}\quad\mbox{and}\quad
 \theta_{l_i} \approx \sqrt{2} {y_i v\over m_\Sigma}
 \end{equation}
respectively.
Due to the $l$--$\Sigma$ mixing (4), we get the mixed gauge
interaction as follows;
 \begin{eqnarray}
&& {\cal L}_{gauge} =
  -g\theta_{\nu_i} W_\mu^+ \left[ {1\over\sqrt{2}} \bar{\Sigma}^0
  \gamma^\mu P_L e_i + \bar{\nu}_i \gamma^\mu R_R \Sigma^- \right]
  \nonumber\\\
   && -g\theta_{\nu_i} W_\mu^- \left[ {1\over\sqrt{2}} \bar{e}_i
 \gamma^\mu P_L \Sigma^0 + \bar{\Sigma}^- \gamma^\mu P_R \nu_i
  \right]\\
 &&+ {g \theta_{\nu_i}\over 2 c_W}  Z_\mu
  \left[ \sqrt{2} \bar{\Sigma}^- \gamma_\mu P_L e_i + \sqrt{2}
  \bar{e}_i \gamma^\mu P_L \Sigma^- - \bar{\Sigma}^0 \gamma^\mu
  \gamma_5 \nu_i \nonumber \right].
 \end{eqnarray}
Thus, the electroweak production of the triplets at the LHC, $pp \to
\Sigma^\pm \Sigma^0, \Sigma^{\pm} \Sigma^{\mp}$, will leave bunch of multi-lepton final
states followed by the triplet decays:
 \begin{eqnarray}
  \Sigma^\pm  &\to& l^\pm h,\quad l^\pm Z^0, \quad \nu W^\pm, \quad \Sigma^0 \pi^\pm \nonumber\\
  \Sigma^0 &\to& \nu h, \quad \nu Z^0, \quad l^\pm W^\mp \,.
 \end{eqnarray}
Among them, trilepton ($3l$) and same sign dilepton (SSD) signals
were shown to provide the most promising
channels for the triplet search \cite{Aguila08}.  This is also true for probing the Higgs boson
in the type III seesaw. That is, the main channels for the Higgs search studied
in this paper will be $3l$ and SSD final states coming from $\Sigma^\pm \Sigma^0$ as follows:
$$
 l^\pm h\, l W(l\nu),\; l^\pm h\, \nu Z(ll),\; l^\pm Z(ll)\, \nu h;\;
 l^\pm h\, l^\pm W^\mp (l\nu,jj) \,.
$$

\begin{table}
\begin{center}
\renewcommand{\arraystretch}{1.0}
\begin{tabular}{|c|c|c|}
\hline
\multicolumn{3}{|c|}{Production cross-sections (fb) } \\
\hline
$m_{\Sigma}$     &~ 250 GeV ~ & ~400 GeV~ \\
\hline\hline
$\Sigma^+\Sigma^0$&439.1&73.8 \\
\hline
$\Sigma^+\Sigma^-$&320.0&50.0 \\
\hline
$\Sigma^-\Sigma^0$&221.8&32.3 \\
\hline
\end{tabular}
\caption{Two benchmark production cross-sections.
} \label{prodcross}
\end{center}
\vspace{-3ex}
\end{table}
\begin{table}
\begin{center}
\renewcommand{\arraystretch}{1.0}
\begin{tabular}{|c|c|c|}
\hline
Decay modes &\multicolumn{2}{|c|}{Branching ratios} \\
\hline
$m_\Sigma$           & 250 GeV  & 400 GeV \\
\hline\hline
$\Sigma^0\to h\nu $&0.17&0.22 \\
\hline
$\Sigma^0\to Z\nu $&0.27&0.26\\
\hline
$\Sigma^0\to W^\pm \l^\mp $&0.56&0.52 \\
\hline
\hline
$\Sigma^\pm\to h l^\pm $&0.17 &0.22\\
\hline
$\Sigma^\pm\to Z l^\pm $&0.27&0.26 \\
\hline
$\Sigma^\pm\to W^\pm\nu $&0.55 &0.52\\
\hline
$\Sigma^\pm\to \Sigma^0\pi^\pm $&0.009&0.003\\
\hline
\end{tabular}
\caption{Triplet branching ratios for $\tilde m_\nu = 10$
meV.}\label{brsigma}
\end{center}
\vspace{-3ex}
\end{table}

For the collider analysis, we have chosen two benchmark points,
BP1 and BP2, with  $m_\Sigma=250$ and 400 GeV, respectively,
taking the Higgs mass of 120 GeV.  The production cross-sections
of the triplet pairs  at the 14 TeV LHC corresponding to BP1 and
BP2 are listed in Table \ref{prodcross}. The branching ratios of
the triplet decay are calculated in Table \ref{brsigma}. The decay
rate of $\Sigma^\pm \to \Sigma^0 \pi^\pm$ is suppressed by a small
mass splitting between $\Sigma^\pm$ and $\Sigma^0$ arising from
one-loop correction, while the other decay rates are proportional
to $y_i^2$ coming from the mixing (4) \cite{Hambye08}. The size of
the neutrino Yukawa coupling $y$ can be quantified by the
effective neutrino mass $\tilde m_\nu \equiv |y|^2 v^2/m_\Sigma$
where we take, in our analysis, $y=y_1$ or $y_2$ denoting the
electron or muon neutrino Yukawa coupling, respectively. When
$\tilde m_\nu$  is sufficiently small, the triplet
decays will occur at large displaced vertices and thus enable us
to trace a displaced Higgs production free from backgrounds
\cite{Bandyo10}. In this work, we do not look for the signatures with displaced
vertices as our analysis can be applied to any value of
$\tilde m_\nu$. For our presentation, we set $\tilde m_\nu=10$ meV
corresponding to the solar neutrino mass scale.

In this study, {\tt MadGraph} \cite{madgraph} has been used for
generating parton-level events for the relevant processes. The
LHEF interface \cite{lhef} was then used to pass the {\tt
MadGraph}-generated events to {\tt PYTHIA}
\cite{Sjostrand:2001yu}. We use {\tt CTEQ6L} parton distribution
function (PDF) \cite{Lai:1999wy,Pumplin:2002vw}. In {\tt MadGraph}
we opted for the lowest order $\alpha_s$ evaluation, which is
appropriate for a lowest order PDF like {\tt CTEQ6L}. The
renormalization/factorization scale in {\tt MadGraph} is set at
$\sqrt{\hat{s}}$. This choice of scale results in a somewhat
conservative estimate for the event rates. ISR/FSR were switched
on in {\tt PYTHIA} for a realistic simulation. For this analysis
we have assumed a $b$-jet tagging efficiency of $\geq$ 50\%
\cite{Baer:2007ya}. For hadronic level simulation we have used
{\tt PYCELL}, the toy calorimeter simulation provided in {\tt
PYTHIA}, with the following criteria: \vspace{-1ex}
\begin{itemize}
  \item the calorimeter coverage is $\rm |\eta| < 4.5$ and the segmentation is
given by $\rm\Delta\eta\times\Delta\phi= 0.09 \times 0.09 $ which resembles
        a generic LHC detector; \vspace{-1ex}
  \item a cone algorithm with
        $\Delta R = \sqrt{\Delta\eta^{2}+\Delta\phi^{2}} = 0.5$
        has been used for jet finding; \vspace{-1ex}
  \item $ p_{T,min}^{jet} = 20$ GeV and jets are ordered in
  $p_{T}$; \vspace{-1ex}
  \item leptons ($\rm \ell=e,~\mu$) are selected with
        $p_T \ge 20$ GeV and $\rm |\eta| \le 2.5$; \vspace{-1ex}
  \item no jet should match with a hard lepton in the event.
\end{itemize}

\begin{figure}
\begin{center}
{\epsfig{file=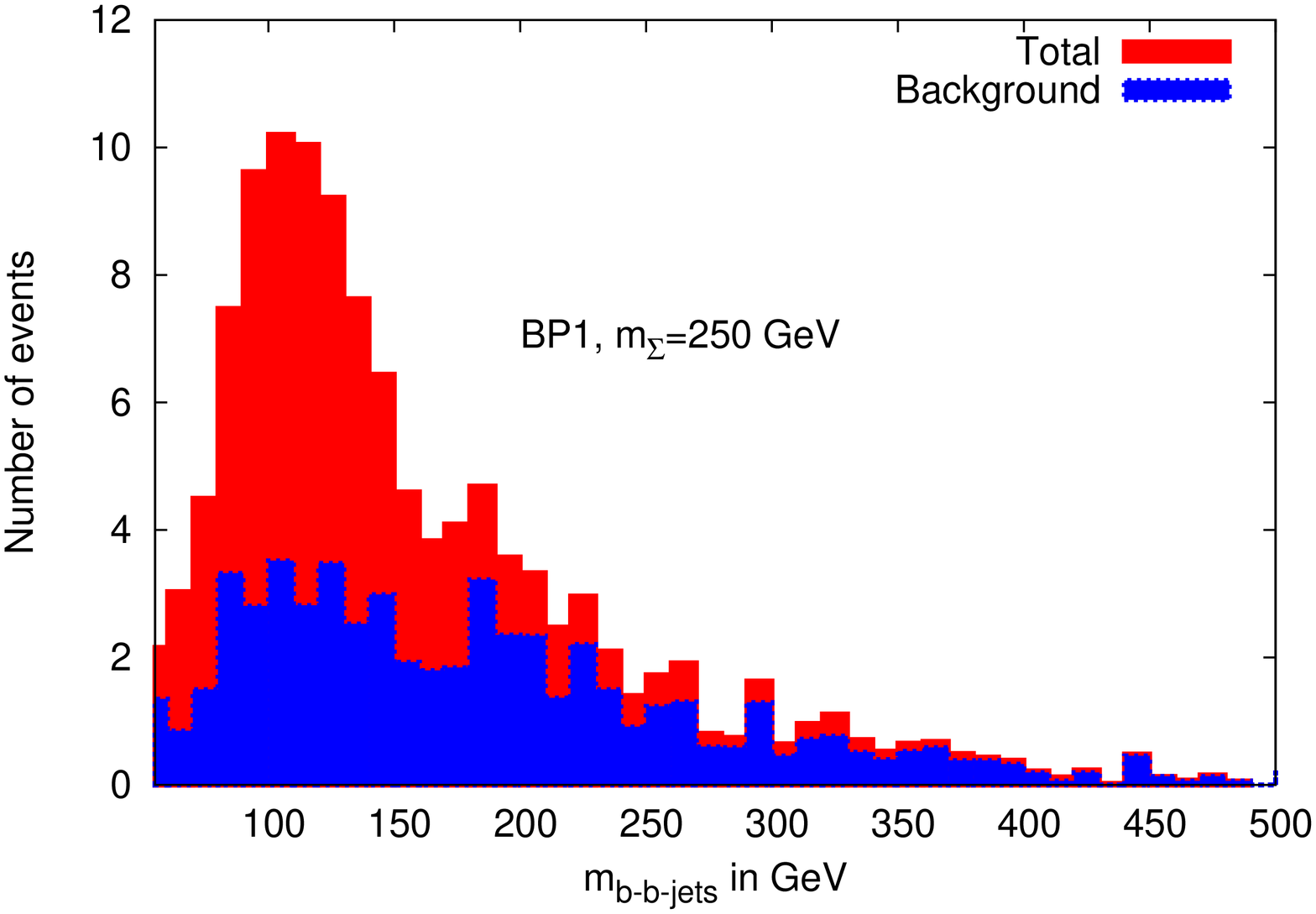,width=4.5cm,height=5.5cm,angle=-90}}
{\epsfig{file=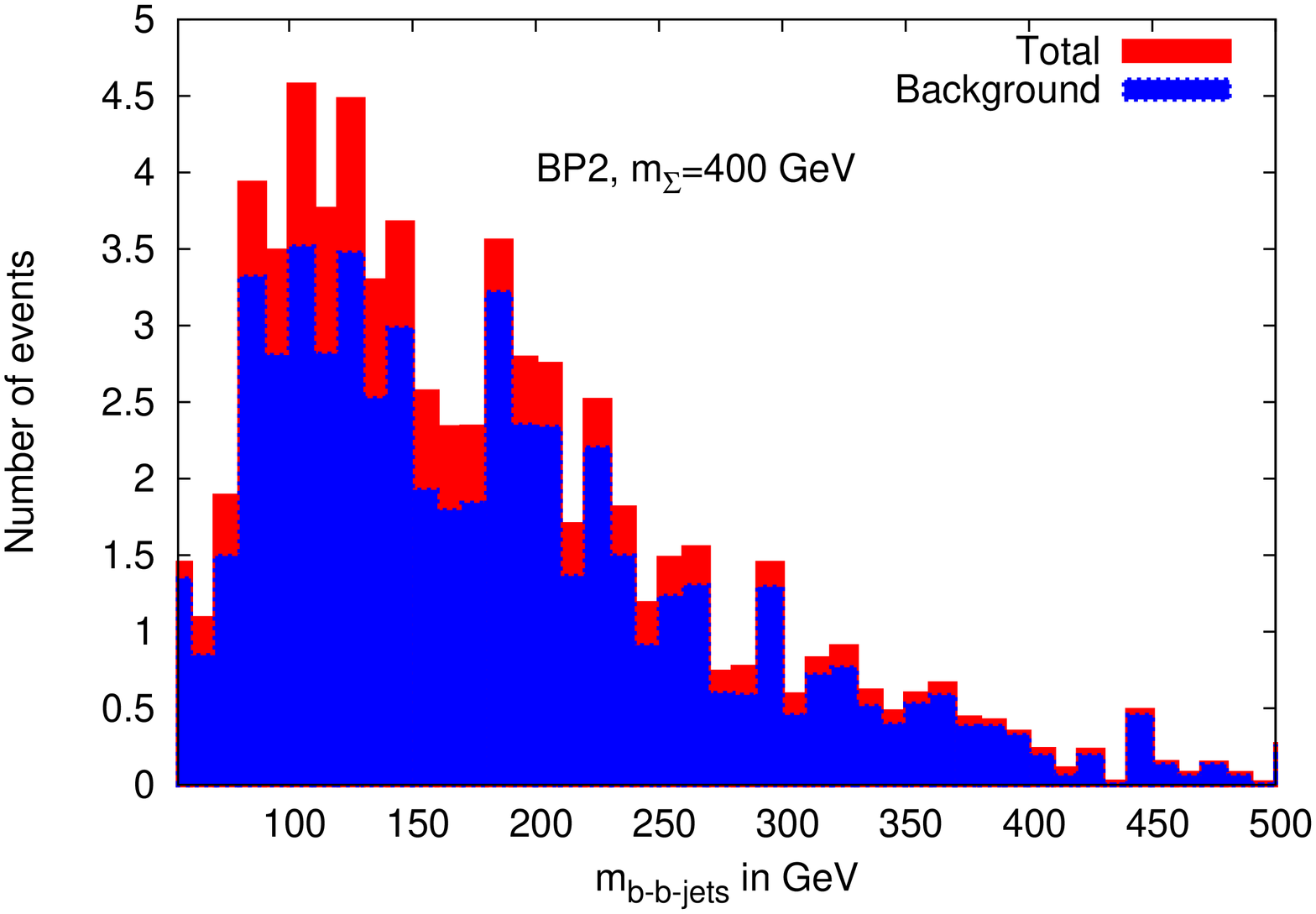,width=4.5cm,height=5.5cm,angle=-90}}
\caption{The $b$--$b$ invariant mass  from $\geq$$ 2b$  $+$ $3l$
final states.
}\label{inH2b3l}
\end{center}
\vspace{-2ex}
\end{figure}
\begin{figure}
\begin{center}
{\epsfig{file=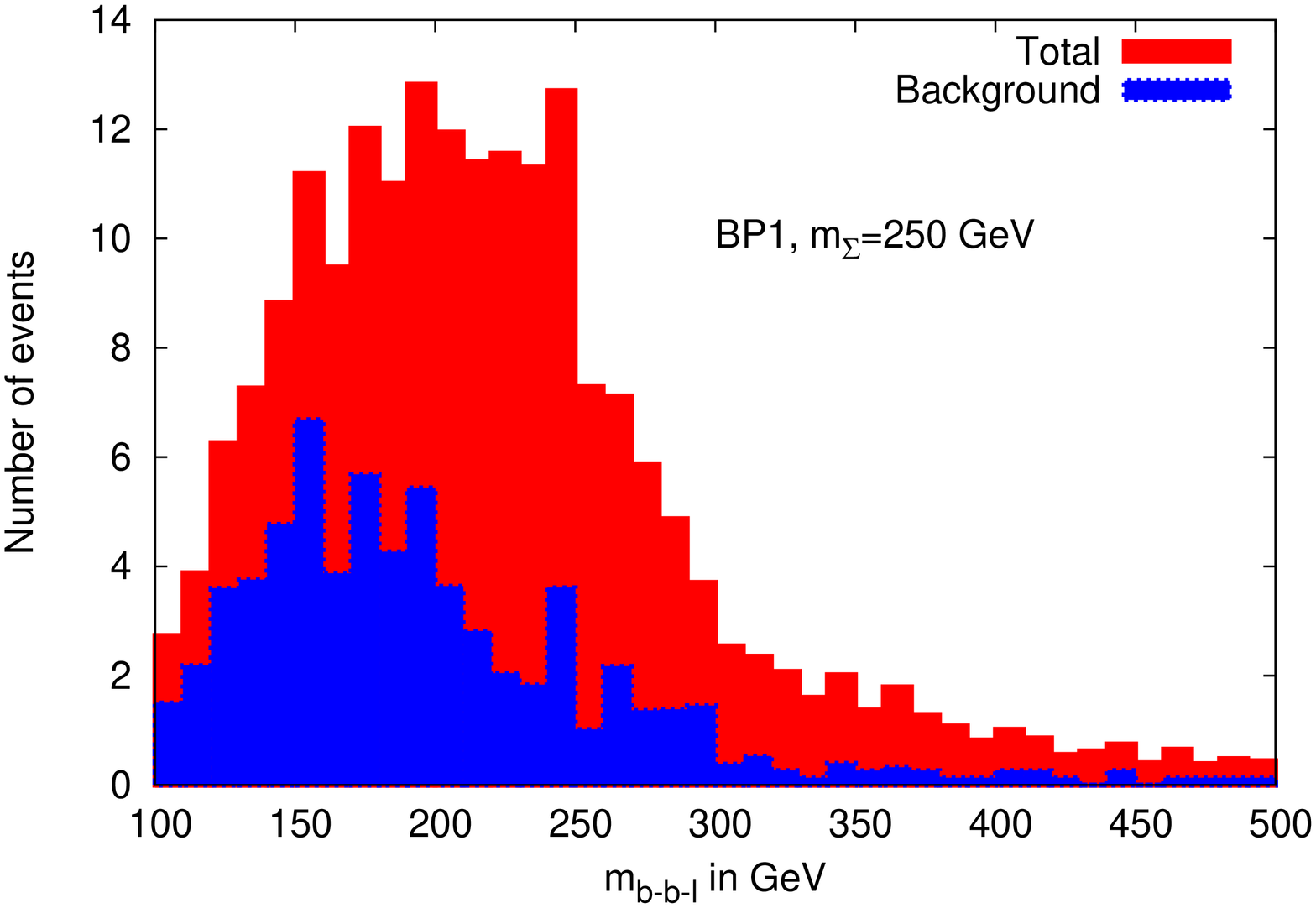,width=4.5cm,height=5.5cm,angle=-90}}
{\epsfig{file=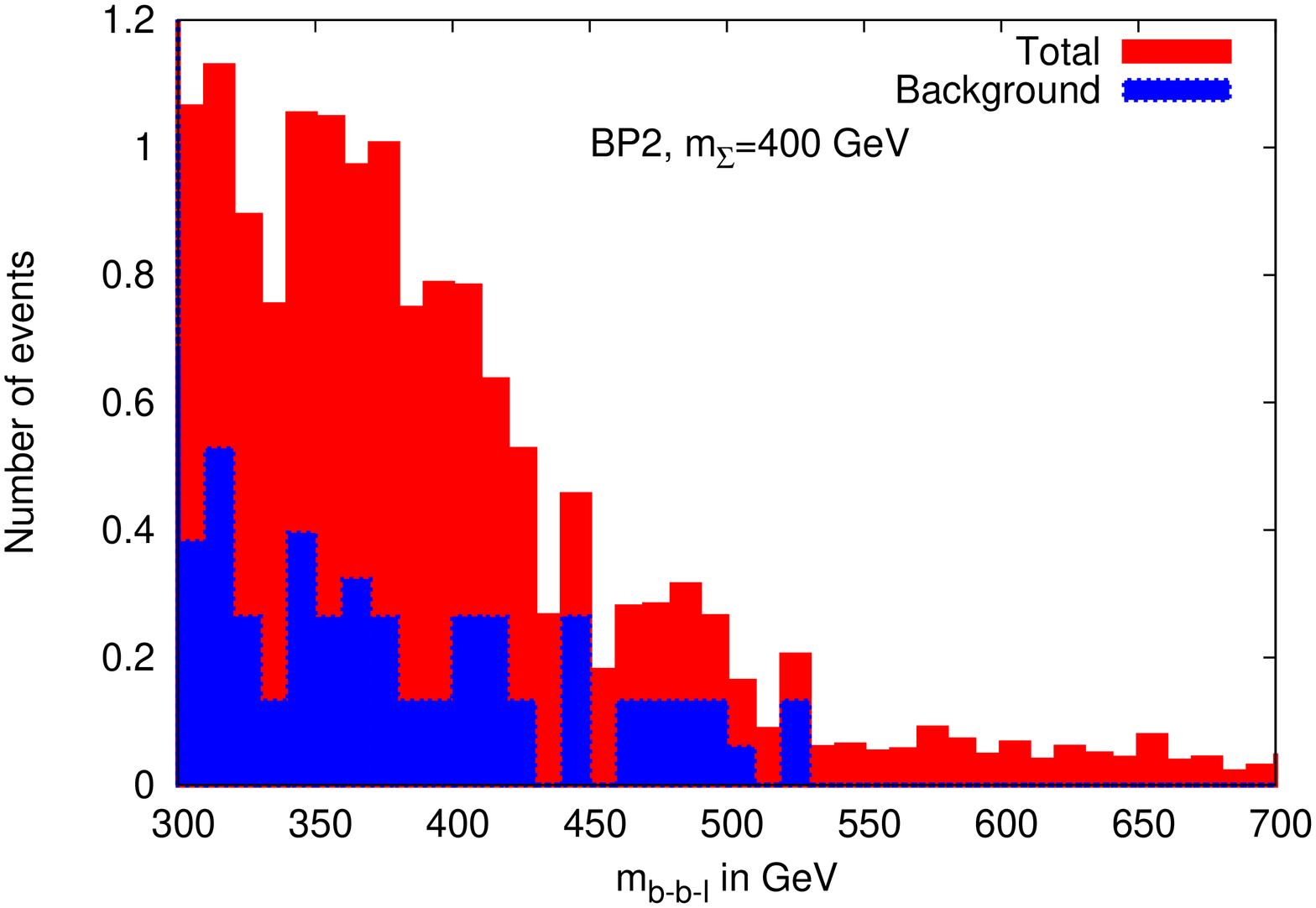,width=4.5cm,height=5.5cm,angle=-90}}
\caption{The $b$--$b$--$l$ invariant mass  from $\geq$$2b$ $+$
$3l$ final states.
}\label{inS2b3l}
\end{center}
\vspace{-3ex}
\end{figure}

\begin{table}
\begin{center}
\renewcommand{\arraystretch}{1.4}
\begin{tabular}{|c|c|c|c|c|c|c|c|}
\hline
 \multicolumn{8}{|c|}{$2 b +  3l$} \\
\hline
\multicolumn{2}{|c|}{Signal}&\multicolumn{6}{|c|}{Backgrounds}\\
\hline
BP1&BP2&$t\bar{t}$&$t\bar{t}b\bar{b}$&$t\bar{t}Z$&$t\bar{t}h$&$VV$&$t\bar{t}W$\\
\hline
116.89& 40.32 &5.0&1.77&31.53&9.86&0.0&8.67\\
\hline
\end{tabular}

\begin{tabular}{|c|c|c|c|c|c|c|c|}
\hline
 \multicolumn{8}{|c|}{$m_{b-b}$} \\
\hline
\multicolumn{2}{|c|}{Signal}&\multicolumn{6}{|c|}{Backgrounds}\\
\hline
BP1&BP2& $t\bar{t}$&$t\bar{t}b\bar{b}$&$t\bar{t}Z$&$t\bar{t}h$&$VV$&$t\bar{t}W$\\
\hline
28.44& 4.46 &1.0&0.50&8.3&2.2&0.0&2.5\\
\hline
\end{tabular}
\begin{tabular}{|c|c|c|c|c|c|c|c|}
\hline
 \multicolumn{8}{|c|}{$m_{b-b-l}$}\\
\hline
\multicolumn{2}{|c|}{Signal}&\multicolumn{6}{|c|}{Backgrounds}\\
\hline
&&$t\bar{t}$&$t\bar{t}b\bar{b}$&$t\bar{t}Z$&$t\bar{t}h$&$VV$&$t\bar{t}W$\\
\hline
BP1&61.89&2.0&0.1&5.9&1.5&0.0&0.9\\
\hline
BP2&5.19&0.0&0.0&1.5&0.06&0.0&0.5\\
\hline
\end{tabular}
\caption{Number of events for $2 b + 3l $; for $m_{b-b}$ within
$120\pm 25$ GeV;  and for $m_{b-b-l}$ within $250\, (400) \pm50$
GeV. }\label{sig2c} \label{sig2h} \label{sig2s}
\end{center}
\vspace{-3ex}
\end{table}

{\bf Higgs search with  $\mathbf{2 b} +  \mathbf{3l}$:}
For the Higgs event selection, we first study the final state topology
with at least two tagged $b$-jets and at least three isolated
leptons. Dominant Standard Model (SM) backgrounds are denoted in Table \ref{sig2c}.
Here $VV+n-{\rm{jets}}$ do not contribute as potential backgrounds.
The number of the signal and background events  for the two
benchmark points are listed in Table \ref{sig2c} for an integrated
luminosity of 10 fb$^{-1}$ at the 14 TeV LHC.
 As we can see from the Table \ref{sig2c} the significance for BP1 is $\sim 9\sigma$ and that of BP2 is $4\sigma$ at 10 fb$^{-1}$ of intregated luminosity.

The $b$-jet pair invariant mass distribution, after the event selection, is presented in
Figure \ref{inH2b3l} which shows the smeared distribution for lower invariant mass due to
the  $Z$ peak contribution. Thus reconstructing Higgs
mass does not increase the signal significance.
Selecting events within the
window of $95 \rm{GeV} \leq m_{b-b} \leq 145\rm{GeV}$ as in
Table \ref{sig2h}, we get the required luminosity for
for $5\sigma$ signal significance 13.3 fb$^{-1}$ for BP1
 and 238fb$^{-1}$ for BP2.

{To check if the reconstructed Higgses are indeed from the triplet
decay, invariant mass distribution of the two $b$-jets with one of
the lepton among the tree isolated leptons is constructed. In
principle among the three leptons the right one will peak at the
triplet mass and the others will contribute in the
 combinatorial background.}
 For this, we select the
the $b$-jets within 60-150 GeV of the invariant mass distribution.
Figure \ref{inS2b3l} describes the invariant of distribution of
$b-b-l$. We then select for $200 \rm{GeV} \leq m_{b-b-l} \leq 300
\rm{GeV}$ for BP1, $350 \rm{GeV} \leq m_{b-b-l} \leq 450 \rm{GeV}$
for BP2. { From the Figure \ref{inS2b3l} we can see that the
distribution has a edge at the triplet mass. This is because, one
low $p_T$ lepton, coming from the decay of gauge boson,
contributes in the lower end of the mass spectrum. This makes the
peak asymmetric in nature. }
 From the signal numbers listed in
Table \ref{sig2s}, one finds the significances of $\sim 7.3
\sigma$ for BP1 and $\sim 2\sigma$ for BP2.

\begin{figure}
\begin{center}
{\epsfig{file=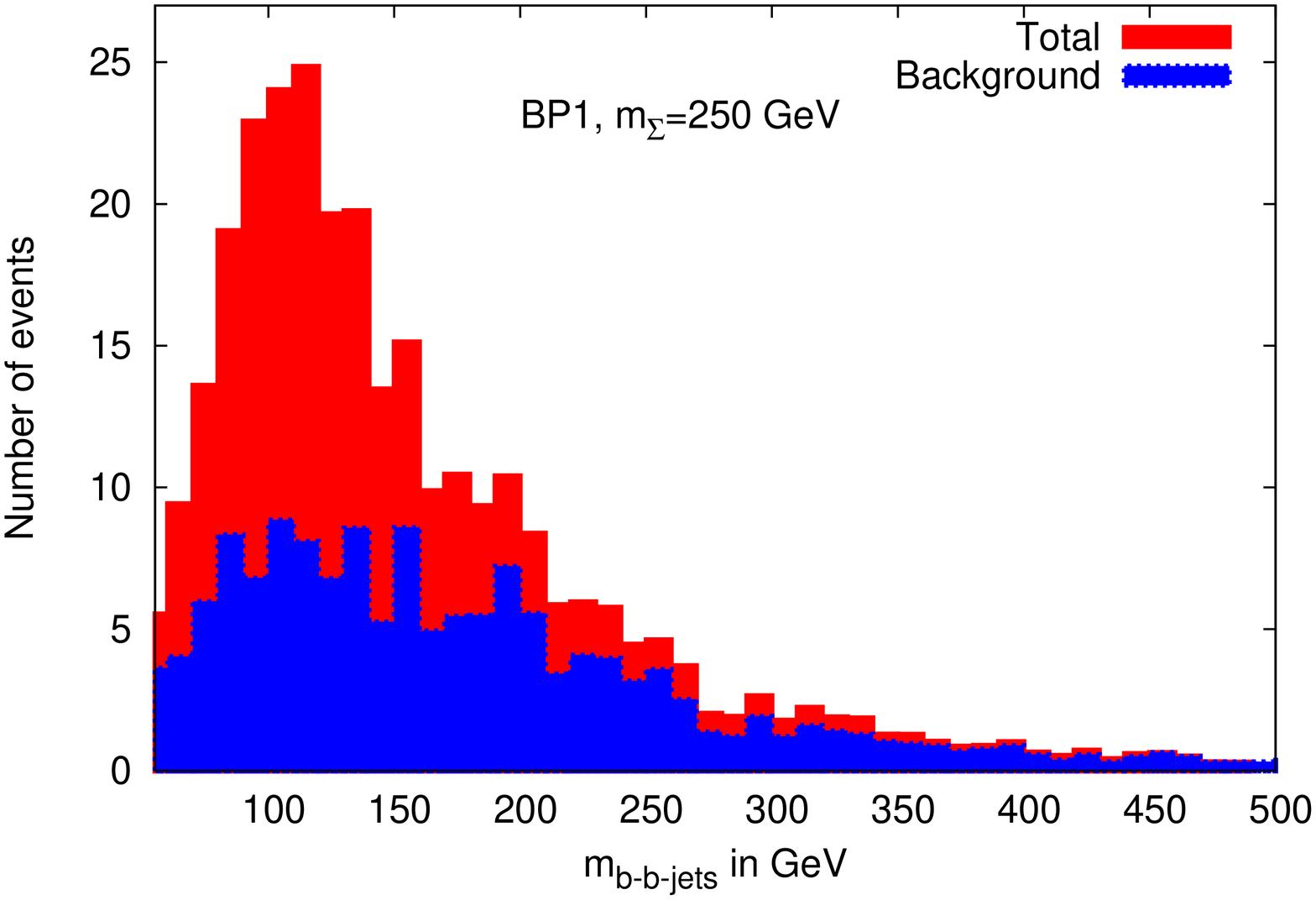,width=4.5cm,height=5.5cm,angle=-90}}
{\epsfig{file=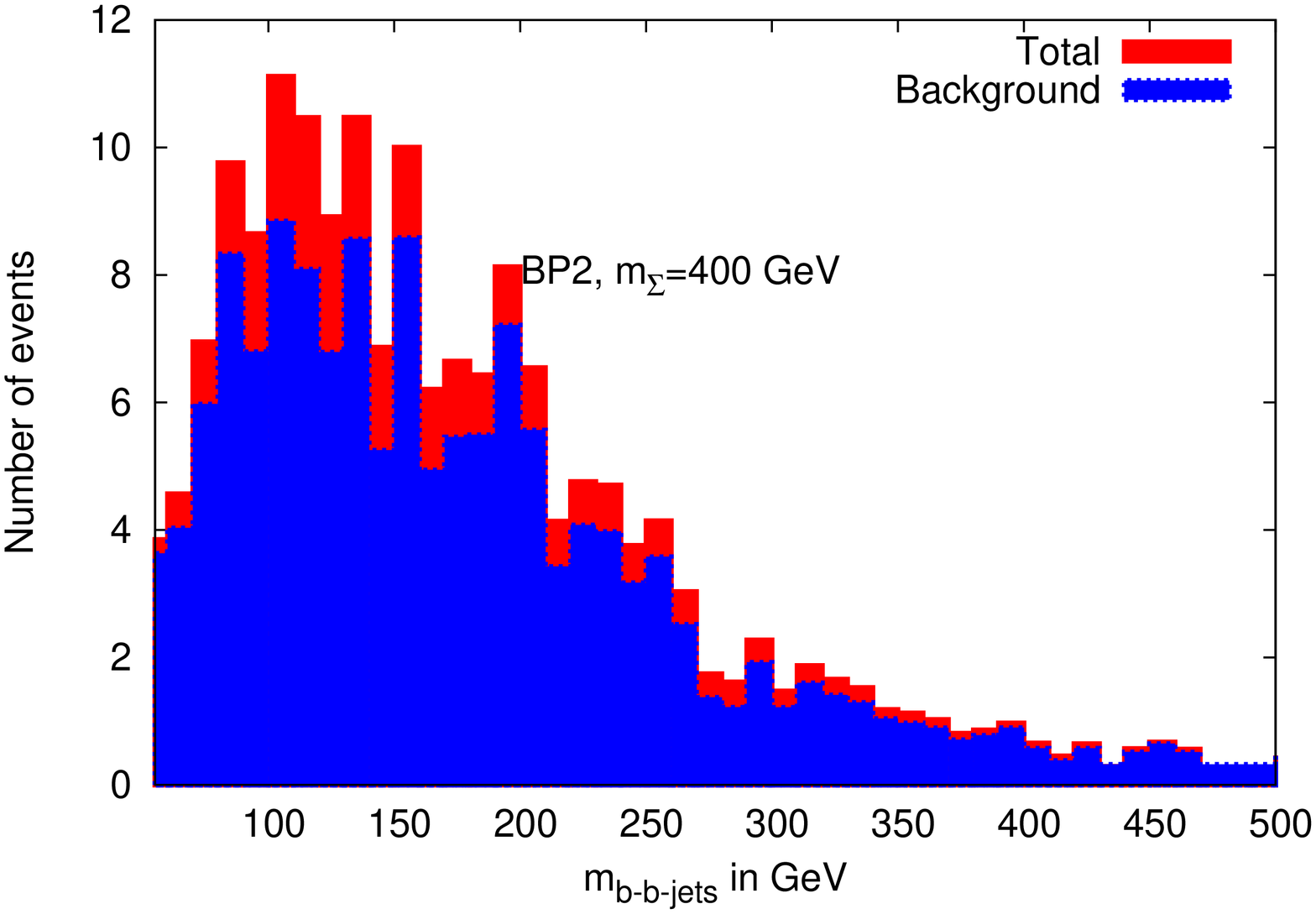,width=4.5cm,height=5.5cm,angle=-90}}
\caption{The $b$--$b$ invariant mass  from $\geq$$ 2b$ $+$ SSD
final states.
}\label{inH2bssd}
\end{center}
\vspace{-3ex}
\end{figure}
\begin{figure}
\begin{center}
{\epsfig{file=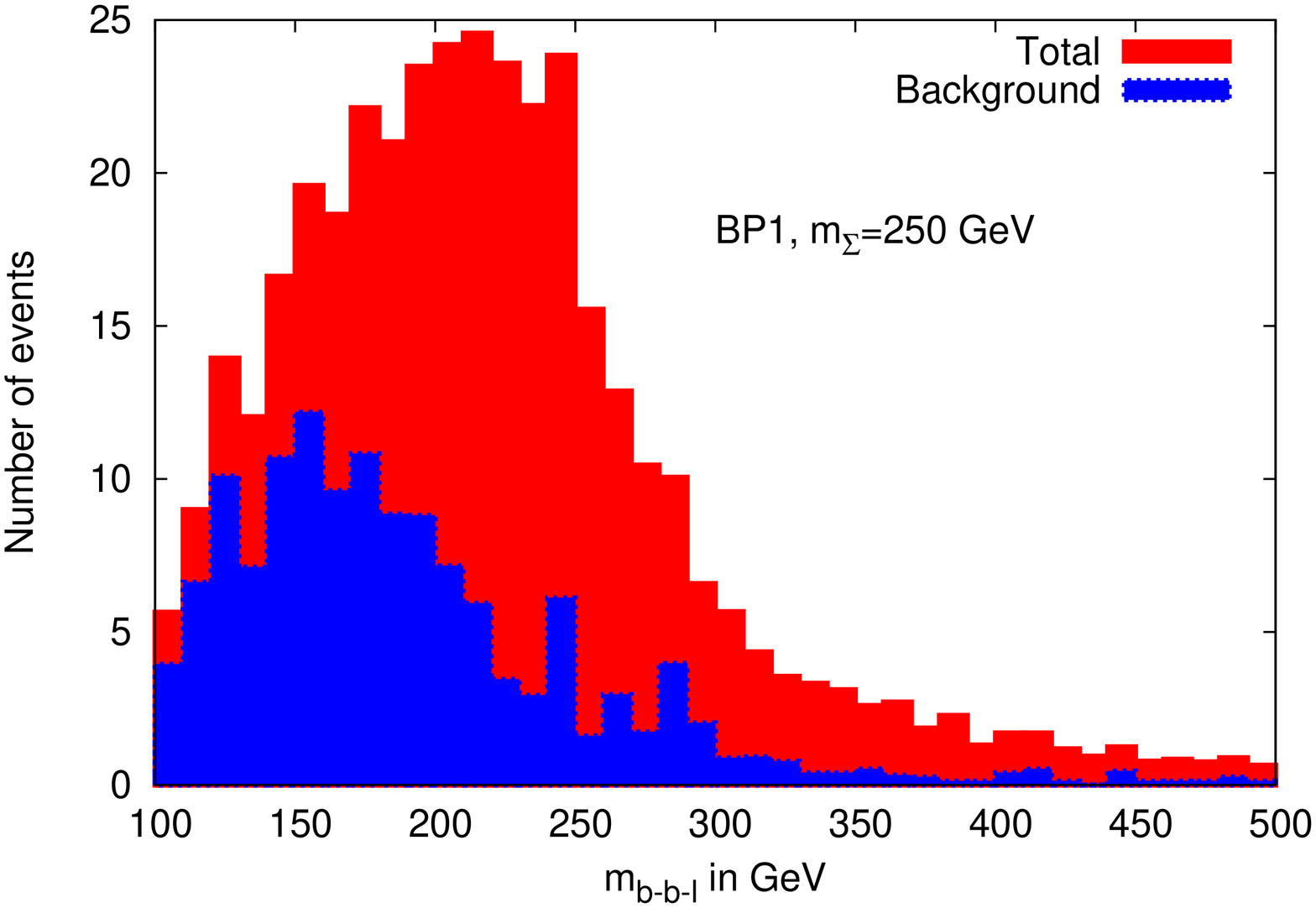,width=4.5cm,height=5.5cm,angle=-90}}
{\epsfig{file=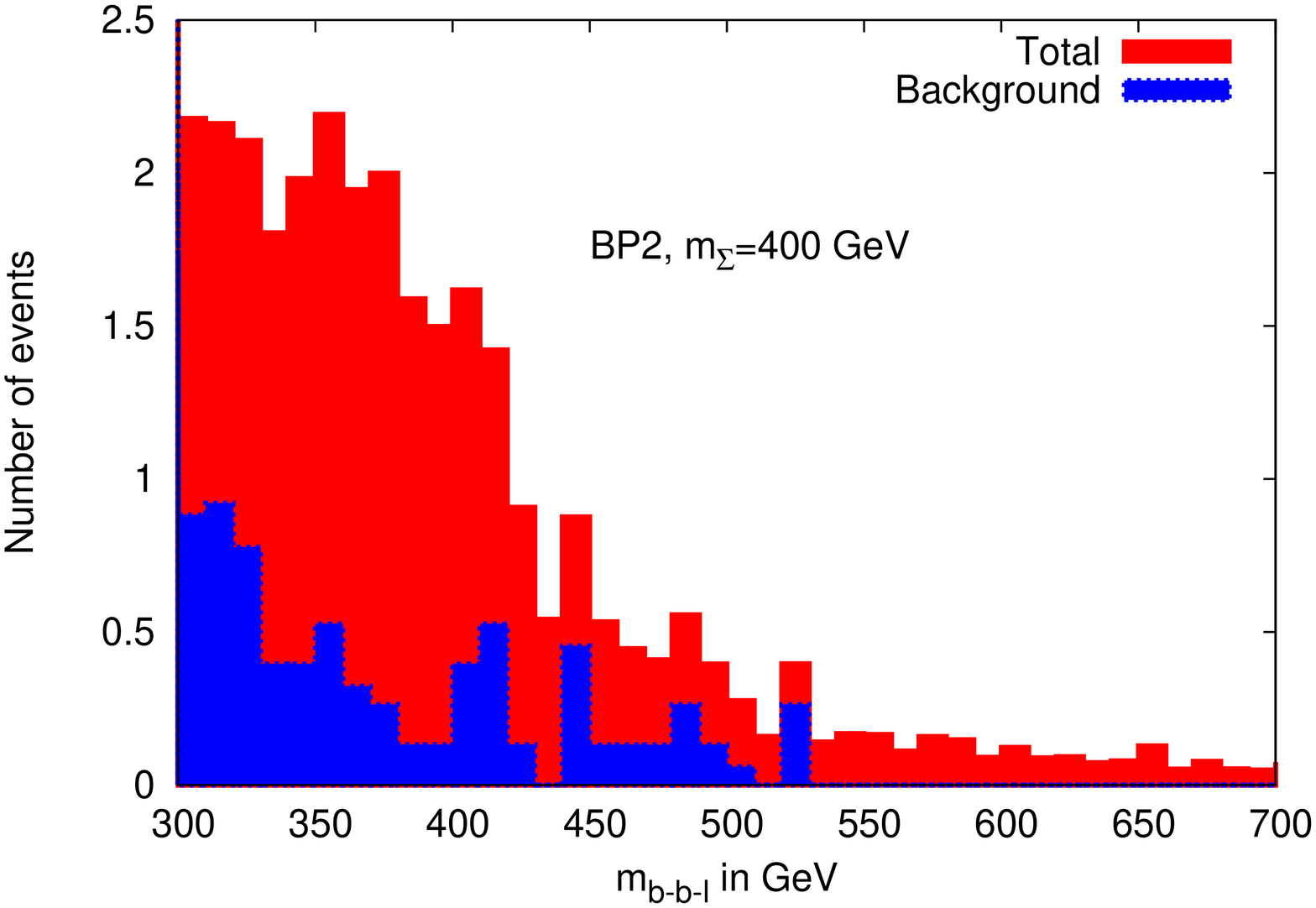,width=4.5cm,height=5.5cm,angle=-90}}
\caption{The $b$--$b$--$l$ invariant mass  from $\geq$$2b$ $+$ SSD
final states.
}\label{inS2bssd}
\end{center}
\vspace{-3ex}
\end{figure}

\begin{table}
\begin{center}
\renewcommand{\arraystretch}{1.4}
\begin{tabular}{|c|c|c|c|c|c|c|c|}
\hline
 \multicolumn{8}{|c|}{$\geq 2 b-\rm{jet}+  SSD$} \\
\hline
\multicolumn{2}{|c|}{Signal}&\multicolumn{6}{|c|}{Backgrounds}\\
\hline
BP1&BP2&$t\bar{t}$&$t\bar{t}b\bar{b}$&$t\bar{t}Z$&$t\bar{t}h$&$VV$&$t\bar{t}W$\\
\hline
127.38& 29.09&24.0&7.5&41.6&29.0&0.0&41.4\\
\hline
\end{tabular}
\begin{tabular}{|c|c|c|c|c|c|c|c|}
\hline
 \multicolumn{8}{|c|}{$m_{b-b}$} \\
\hline
\multicolumn{2}{|c|}{Signal}&\multicolumn{6}{|c|}{Backgrounds}\\
\hline
BP1&BP2 &$t\bar{t}$&$t\bar{t}b\bar{b}$&$t\bar{t}Z$&$t\bar{t}h$&$VV$&$t\bar{t}W$\\
\hline
60.61& 10.39&8.0&3.0&10.6&6.5&0.0&9.6\\
\hline
\end{tabular}
\begin{tabular}{|c|c|c|c|c|c|c|c|}
\hline
 \multicolumn{8}{|c|}{$m_{b-b-l}$}\\
\hline
\multicolumn{2}{|c|}{Signal}&\multicolumn{6}{|c|}{Backgrounds}\\
\hline
&&$t\bar{t}$&$t\bar{t}b\bar{b}$&$t\bar{t}Z$&$t\bar{t}h$&$VV$&$t\bar{t}W$\\
\hline
BP1&117.37&4.0&0.35&7.00&2.67&0.0&2.50\\
\hline
BP2&11.74&0.0&0.0&1.71&0.12&0.0&1.05\\
\hline
\end{tabular}
\caption{Number of events for $2 b + \mbox{SSD}$; for $m_{b-b}$
within $120\pm 25$ GeV;  and for $m_{b-b-l}$ within $250\, (400)
\pm50$ GeV.} \label{sig4h} \label{sig4c} \label{sig4s}
\end{center}
\vspace{-3ex}
\end{table}

{\bf Higgs search with $\mathbf{2b} + \mbox{SSD}$: }
The final states with same-sign dileptons are also free from
severe Standard Model backgrounds.
The final decay modes from $\Sigma$ that contribute to the process are
$hlWl$, $hlZl$, $ZlZl$, $ZlWl$ respectively. Basically, the
$3l$ events always carry same sign dileptons  and will
contribute to this final state. In addition, signal acceptance
is gained by requiring two or more leptons.
If one of the leptons in a trilepton event does not pass the acceptance,
there is a chance that it will still be accepted in the SSD selection.

We select the events with at least two $b$-tagged jets and at
least two isolated same-sign leptons in the final states. In Table
\ref{sig4c} we present the corresponding signal numbers  for the
two benchmark points and the SM backgrounds at 10 fb$^{-1}$ of
integrated luminosity. The signal significances are $\sim 8\sigma$
for BP1, and  $2.2\sigma$  for BP2.

After analyzing the final
state with SSD and $2b$-jets we construct the Higgs mass
peak. We plot the invariant mass distribution of these to $b$-jets
which peak around the $Z$ and Higgs mass  in Figure \ref{inH2bssd}.
Selecting  events $95 \rm{GeV} \leq m_{b-b} \leq 145\rm{GeV}$ we
get the signal numbers for the Higgs mass peak as listed in Table \ref{sig4h}.
The corresponding signal significance over SM backgrounds
are $6.11\sigma$, and $1.5\sigma$ for BP1, and BP2, respectively.
This shows the SSD events are better than the $3l$ events in probing the Higgs boson.

{ Similar to the $3l$ case, we reconstruct the triplet mass from
the selected Higgs events and the leptons in the final state. For
that we take again the $b$-jets within the mass window of 60-150
GeV and plot the invariant mass distribution with the lepton in
the final state in Figure \ref{inS2bssd}.} Selecting the events
within the windows of $200 \rm{GeV} \leq m_{b-b-l} \leq 300
\rm{GeV}$
 for BP1, and  $350 \rm{GeV} \leq m_{b-b-l} \leq 450 \rm{GeV}$
 for BP2, we get the result in Table \ref{sig4s}.
In this case the significance really gets enhanced;
$\simeq 10\sigma$ for BP1, and $3\sigma$ for BP2.

\smallskip

{In conclusion, we examined $b$-jet pair signals from Higgs in association with
trileptons or same-sign dileptons in the type III seesaw mechanism.
These channels enough significances over the Standard Model
backgrounds, in particular, in the high $p_T$ regime, and thus
provide viable channels for the light Higgs boson search. Early
data of the 14 TeV LHC will enable us to reconstruct the Higgs
boson coming from the triplet decay through  $b$--$b$ and
$b$--$b$--$l$ invariant mass distributions for relatively lower
triplet mass. On the other hand the reach  goes down rapidly for
higher triplet mass due to the strong depletion of the triplet
production cross-section which will be overcome by further
accumulation of the luminosity.

\smallskip
{\bf Acknowledgments:}
 EJC was supported by Korea Neutrino
Research Center through National Research Foundation Grant
(2009-0083526).
 SC was supported by National
Research Foundation Grant (2009-0069251). We also acknowledge
Roberto Francheschini for providing us with the MadGraph model
files implementing the Type III seesaw model.


\begin{thebibliography}{99}

\bibitem{lhc1213}
ATLAS Collaboration, ATLAS-CONF-2011-163; CMS Collabortion, CMS-PAS-HIG-11-032.

\bibitem{Butterworth08}
  J.~M.~Butterworth, A.~R.~Davison, M.~Rubin and G.~P.~Salam,
  Phys.\ Rev.\ Lett.\  {\bf 100}, 242001 (2008)
 [arXiv:0802.2470 [hep-ph]].

\bibitem{Foot88}
  R.~Foot, H.~Lew, X.~G.~He and G.~C.~Joshi,
  Z.\ Phys.\  C {\bf 44} (1989) 441.

\bibitem{Hambye08}
  R.~Franceschini, T.~Hambye and A.~Strumia,
  Phys.\ Rev.\  D {\bf 78} (2008) 033002
  [arXiv:0805.1613 [hep-ph]].

\bibitem{Aguila08}
  F.~del Aguila and J.~A.~Aguilar-Saavedra,
  Nucl.\ Phys.\  B {\bf 813} (2009) 22
  [arXiv:0808.2468 [hep-ph]].

\bibitem{Arhrib09}
  A.~Arhrib {\it et al.},
  [arXiv:0904.2390 [hep-ph]].

\bibitem{Bandyo09}
  P.~Bandyopadhyay, S.~Choubey and M.~Mitra,
  JHEP {\bf 0910} (2009) 012
  [arXiv:0906.5330 [hep-ph]].


\bibitem{Li09}
  T.~Li and X.~G.~He,
  Phys.\ Rev.\  D {\bf 80} (2009) 093003
  [arXiv:0907.4193 [hep-ph]].



\bibitem{Bandyo10}
  P.~Bandyopadhyay and E.~J.~Chun,
  JHEP {\bf 1011}, 006 (2010)
  [arXiv:1007.2281 [hep-ph]].

\bibitem{madgraph}
  J.~Alwall
  {\it et al.},
  JHEP {\bf 0709}, 028 (2007)
  [arXiv:0706.2334 [hep-ph]].


\bibitem{lhef}
  J.~Alwall {\it et al.},
  Comput.\ Phys.\ Commun.\  {\bf 176}, 300 (2007)
 [arXiv:hep-ph/0609017].


\bibitem{Sjostrand:2001yu}
  T.~Sjostrand, L.~Lonnblad and S.~Mrenna,
  [arXiv:hep-ph/0108264].

\bibitem{Lai:1999wy}
  H.~L.~Lai {\it et al.}  [CTEQ Collaboration],
  Eur.\ Phys.\ J.\ C {\bf 12}, 375 (2000)
  [arXiv:hep-ph/9903282].

\bibitem{Pumplin:2002vw}
  J.~Pumplin {\it et al.}, 
  JHEP {\bf 0207}, 012 (2002)
  [arXiv:hep-ph/0201195].

\bibitem{Baer:2007ya}
  H.~Baer {\it et al.}, 
  Phys.\ Rev.\  D {\bf 75}, 095010 (2007)
  [arXiv:hep-ph/0703289].

\end{thebibliography}
\end{document}